\documentclass{elsart}
\usepackage{graphicx,amssymb}
\usepackage{epsfig}

\newcommand{\pict}{pict}

\renewcommand{\textfraction}{.3}
\renewcommand{\floatpagefraction}{.7}

\begin{document}

\begin{frontmatter}

\title{The Outer Tracker Detector of the HERA-B Experiment\\ 
Part\,II: Front-End Electronics \\[1cm]
{\large HERA-B Outer Tracker Group}}
%
%
%
%
%
\author[hamburg]{H.~Albrecht},  
\author[hamburg]{M.~Bahte}, 
\author[nikhef,utrecht]{Th.~S.~Bauer}, 
\author[rostock]{M.~Beck},   
\author[zeuthen]{K.~Berkhan},  
\author[zeuthen]{G.~Bohm},  
\author[nikhef,utrecht]{M.~Bruinsma}, 
\author[oslo]{T.~Buran},
\author[hamburg]{M.~Capeans}, 
\author[tsinghua]{B.~X.~Chen}, 
\author[berlin]{H.~Deckers}, 
\author[ihep]{X.~Dong},  
\author[austin]{R.~Eckmann}, 
\author[hamburg]{D.~Emelianov},  
\author[hamburg,mephi]{G.~Evgrafov},  
\author[dubna]{I.~Golutvin},  
\author[hamburg]{M.~Hohlmann},
\author[hamburg]{K.~H\"opfner}, 
\author[nikhef]{W.~Hulsbergen},  
\author[ihep]{Y.~Jia},  
\author[ihep]{C.~Jiang}, 
\author[dortmund]{H.~Kapitza}, 
\author[hamburg]{S.~Karabekyan},  
\author[ihep]{Z.~Ke},  
\author[dubna]{Y.~Kiryushin},
\author[berlin]{H.~Kolanoski\corauthref{cor}},
\corauth[cor]{Corresponding address: DESY, Platanenallee 6, 
D-15738 Zeuthen, Germany, Tel.: +49-33762-77380, fax: +49-33762-77330. }
\ead{hermann.kolanoski@desy.de}
\author[stefan,maribor]{S.~Korpar},  
\author[stefan,ljubljana]{P.~Kri\v zan},  
\author[berlin]{D.~Kr\"ucker}, 
\author[dubna]{A.~Lanyov}, 
\author[tsinghua]{Y.~Q.~Liu}, 
\author[berlin]{T.~Lohse},   
\author[berlin]{R.~Loke},   
\author[berlin]{R.~Mankel}, 
\author[berlin]{G.~Medin\thanksref{gk}},  
\author[hamburg]{E.~Michel},
\author[dubna]{A.~Moshkin}, 
\author[tsinghua]{J.~Ni},   
\author[zeuthen]{S.~Nowak}, 
\author[nikhef,utrecht]{M.~Ouchrif}, 
\author[hamburg]{C.~Padilla},
\author[rostock]{R.~Pernack}, 
\author[hamburg,itep]{A.~Petrukhin}, 
\author[zeuthen]{M.~Pohl}, 
\author[dubna,heidelberg]{D.~Pose},  
\author[hamburg]{D.~Ressing},
\author[hamburg]{B.~Schmidt},
\author[unihh]{W.~Schmidt-Parzefall}, 
\author[zeuthen]{A.~Schreiner},  
\author[hamburg,rostock]{H.~Schr\"oder}, 
\author[zeuthen]{U.~Schwanke}, 
\author[hamburg]{A.~S.~Schwarz},  
\author[hamburg]{I.~Siccama}, 
\author[dubna]{S.~Solunin},
\author[hamburg]{S.~Somov}, 
\author[zeuthen]{V.~Souvorov},  
\author[zeuthen,itep]{A.~Spiridonov},
\author[stefan]{M.~Stari\v c},  
\author[zeuthen,berlin]{C.~Stegmann}, 
\author[nikhef]{O.~Steinkamp}, 
\author[hamburg]{N.~Tesch}, 
\author[hamburg,sofia]{I.~Tsakov},  
\author[berlin,heidelberg]{U.~Uwer},  
\author[dubna]{S.~Vassiliev}, 
\author[dubna]{D.~Vishnevsky},
\author[berlin,zeuthen]{I.~Vukotic\thanksref{gk}}, 
\author[zeuthen]{M.~Walter},  
\author[tsinghua]{J.~J.~Wang},
\author[tsinghua]{Y.~M.~Wang},
\author[zeuthen]{H.~J.~Wilczek},
\author[hamburg]{R.~Wurth},   
\author[tsinghua]{J.~Yang},   
\author[ihep]{Z.~Zheng}, 
\author[ihep]{Z.~Zhu},   
\author[rostock]{R.~Zimmermann}

\address[nikhef]{NIKHEF, Kruislaan 409, PO Box 41882, 1009 DB Amsterdam, 
Netherlands\thanksref{nl}}

\address[austin]{University of Texas at Austin, Department of Physics, RLM
  5.208, Austin TX 78712-1081, USA\thanksref{usa}}

\address[ihep]{Institute of High Energy Physics, Beijing 100039, China}
\address[tsinghua]{Institute of Engineering Physics, Tsinghua University,
  Beijing 100084 , P.R. China}

\address[berlin]{Institut f\"ur Physik, Humboldt Universit\"at zu Berlin,
  D-12489 Berlin, Germany\thanksref{bmbf}}

\address[dortmund]{Institut f\"ur Physik, Universit\"at Dortmund, D-4427
  Dortmund, Germany\thanksref{bmbf}} 

\address[dubna]{Joint Institute for Nuclear Research, Dubna, RU-141980, Russia}
\address[hamburg]{DESY, D-22607 Hamburg, Germany}
\address[unihh]{Institut f\"ur Experimentalphysik, Universit\"at Hamburg,
  D-22761 Hamburg, Germany\thanksref{bmbf}}

\address[heidelberg]{Physikalisches Institut, Universit\"at Heidelberg,
  D-69120 Heidelberg, Germany\thanksref{bmbf}}

\address[stefan]{Jozef Stefan Institute, 1001 Ljubljana, Slovenia}
\address[ljubljana]{University of Ljubljana, 1001 Ljubljana, Slovenia}
\address[maribor]{University of Maribor, 2000 Maribor, Slovenia} 
\address[itep]{Institute of Theoretical and Experimental Physics, 117259
  Moscow, Russia}        

\address[oslo]{Institute of Physics, University of Oslo,
  Norway\thanksref{norway}}

\address[rostock]{Fachbereich Physik, Universit\"at Rostock, D-18051 Rostock,
  Germany\thanksref{bmbf}}

\address[utrecht]{Universiteit Utrecht/NIKHEF, 3584 CB Utrecht, The Netherlands\thanksref{nl}}

\address[zeuthen]{DESY, D-15738 Zeuthen, Germany}

\address[mephi]{visitor from Moscow Physical Engineering Institute, 115409
  Moscow, Russia}

\address[sofia]{visitor from Institute for Nuclear Research, INRNE-BAS, Sofia,
  Bulgaria}

\address[yerevan]{visitor from Yerevan Physics Institute, Yerevan, Armenia}

\thanks[nl]{Supported by the Foundation for Fundamental Research on Matter
(FOM), 3502 GA Utrecht, The Netherlands}

\thanks[usa]{Supported by the U.S. Department of Energy (DOE)}

\thanks[bmbf]{Supported by Bundesministerium f\"ur Bildung und Forschung,
  FRG, under contract numbers  05-7BU35I, 05-7D055P, 05 HB1KHA, 05 HB1HRA,
  05 HB9HRA, 05 7HD15I, 05 7HH25I} 


\thanks[norway]{Supported by the Norwegian Research Council}

\thanks[gk]{Supported by German Research Foundation (Research Training
Group GK 271)}

\begin{abstract} 
  The HERA-B Outer Tracker is a large detector with 
  112\,674 drift chamber channels. It is exposed to a particle flux
  of up to $2\cdot 10^5$\,cm$^{-2}$ s$^{-1}$ thus coping with
  conditions similar to those expected for the LHC experiments. The
  front-end readout system, based on the ASD-8 chip
  and a customized TDC chip, is designed
  to fulfil the requirements on  low noise,
  high sensitivity, rate tolerance, and high integration density.
  
  The TDC system is based on an ASIC which digitizes the time in bins
  of about 0.5\,ns within a total of 256 bins. The chip also comprises
  a pipeline to store data from 128 events which is required for a
  deadtime-free trigger and data acquisition system.
  
  We report on the development, installation, and commissioning of the
  front-end electronics, including the grounding and noise suppression
  schemes, and discuss its performance in the HERA-B experiment.

\end{abstract}

\end{frontmatter}

%

\section{Introduction}

HERA-B was designed as a fixed target experiment for studying CP
violation in $B$-meson systems using an internal wire target in the
proton beam of HERA \cite{herab}. To reach the necessary production
rate of $b$ quarks an average of four interactions per bunch crossing at
a frequency of about 10 MHz (96 ns bunch separation) has to be
generated. This leads to a very high particle flux density in the
detector.

The main detector components (fig.\,\ref{figotr}) are a silicon vertex
detector, a dipole magnet with a field integral of 2.13\,Tm, a main
tracker with an Inner Tracker composed of microstrip gas chambers
and an Outer Tracker composed of drift tubes, High-p$_T$ 
Chambers, a ring imaging Cherenkov counter (RICH),
an electromagnetic calorimeter (ECAL), and a Muon
System with drift tubes.  The detector covers a forward angular
range of 220 mrad in the bending plane of the magnet and 160 mrad
vertically.

In the following we describe the front-end electronics of the Outer
Tracker drift tubes, i.\ e.\ the outer part of the main tracking
system, which is based on the amplifier-shaper-discriminator chip
ASD-8 \cite{asdpub} and a TDC (time-to-digital-converter) chip
customized for HERA-B.

The paper is organized as follows: in the next section the Outer
Tracker detector is briefly described followed in section 3 by a
discussion of the design considerations for the front-end
electronics. The sections 4 to 6 contain the description of the main
components of the front-end electronics: the high-voltage system, the
ASD8 board and the TDC board. Section 7 describes the installation and
commissioning of the electronics and gives a first evaluation of the
performance. The paper finishes with a summary.


\section{The Outer Tracker System}
\label{secotr}
\begin{figure}
     \begin{center} \leavevmode
\epsfxsize=\textwidth  \epsfbox{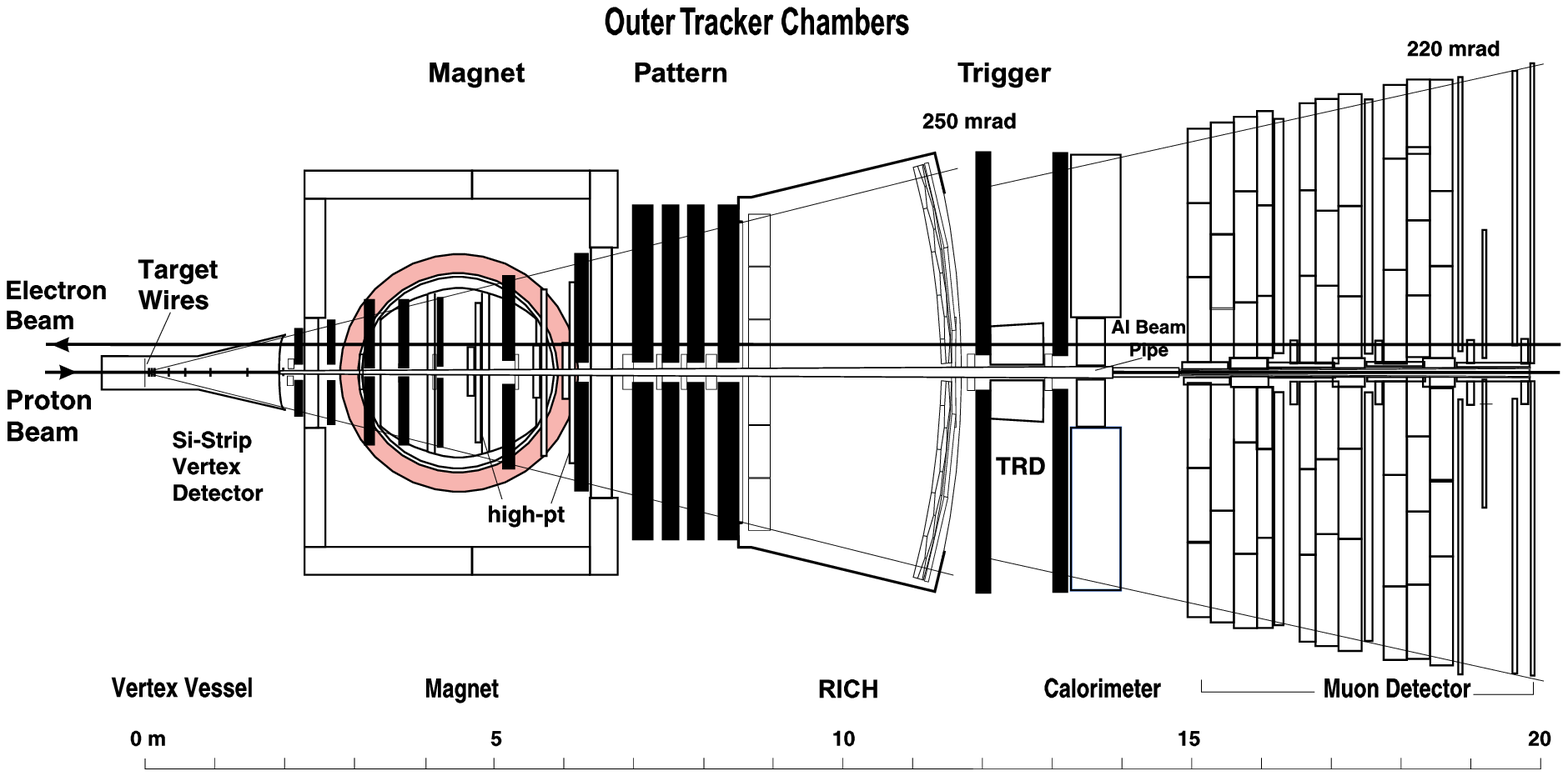}
\end{center} 
\caption{Top view of the HERA-B detector. The Outer Tracker
  superlayers, arranged as magnet, pattern and trigger chambers, are
  indicated by the black areas (with the Inner Tracker modules
  attached to them near the beam pipe, white areas).}
\label{figotr}
\end{figure}  
\paragraph*{Detector Description:}
The Outer Tracker of HERA-B \cite{herab,otr_det} is composed of 13
planar superlayers (fig.\,\ref{figotr}) of drift tube modules
comprising 112\,674 readout channels. Each superlayer consists of two
independent chambers. For each chamber the modules are contained in a
gas-tight box which is suspended from a rigid frame (`outer
frame'). Of the 13 superlayers 7 are placed in the magnet (`magnet
chambers' = MC), 4 in the field-free region between the magnet and the
RICH serving for pattern recognition (`pattern chambers'= PC) and 2
between the RICH and the ECAL (`trigger chambers' = TC).  The two TC
superlayers and the first and last PC superlayers deliver hit signals
to the First Level Trigger. The main tracking system allows for
momentum measurement and provides track recognition on the first
trigger level.

The honeycomb modules of hexagonal drift tubes are built from folded
gold-coated, carbon-loaded polycarbonate foil.  The tubes are oriented
at $0^{\circ}$ and $\pm 5^{\circ}$ relative to the perpendicular on
the bending plane.  In the bending plane the inner diameter of the
cells changes from 5\,mm near the beam to 10 mm above a certain
distance from the beam (about 1\,m in the PC area) to account for the
radial dependence of the particle flux.

As counting gas the fast mixture Ar/CF$_4$/CO$_2$ (65/30/5) is chosen.
Operating at a gain of $3\cdot 10^4$, the drift velocity is about
80\,$\mu$m/ns which allows the particle signals to be collected within
the bunch separation time of 96\,ns, even within the magnetic field.  In
this gas the ionisation density of a minimum ionising track is
estimated to be about 100/cm with an average cluster size of 2.5
electrons. However, due to the loss of electrons by attachment
only about 30\% of the electrons reach the anode, leading
to a mean cluster size of about 1 electron.

\paragraph*{Radiation load and occupancies:}
The Outer Tracker has been designed for particle densities and
radiation levels comparable to those in similar detectors currently
developed for LHC. At an interaction rate of 40\,MHz the radial
distribution of the charged particle flux density is roughly given by:
$$
\phi \approx \frac{10^{8}}{R^2}\,{\rm Hz}
$$ 
where $R$ is the distance from the beam (in any units).  Since the
Outer Tracker acceptance starts at a radial distance of about 19~cm to
the beam this leads to a particle flux of about $2\cdot 10^{5}\,{\rm
cm}^{-2} s^{-1}$ in the hottest area.

The drift tubes are longitudinally subdivided into sensitive and
insensitive segments to limit the single channel occupancy to about
20\% at an interaction rate of 40\,MHz (the shortest segmentation near
the beam is 20 cm). The subdivision is achieved by installing a sense
wire only in the sensitive parts of a cell \cite{otr_det}. If a
sensitive part is not at an end of the cell the signal is carried to
the upper or lower end of the module via a thicker wire which does
not generate an avalanche. The sensitive length of the drift cells,
$L_{cell}$, varies between 0.2\,m and 2.8\,m yielding different cell
capacities. The capacitive load at the input of the amplifier is about
$15\,{\rm pF} + L_{cell} \cdot 10\,{\rm pF/m}$, resulting in loads
between 17 and 43\,pF.

The amplifiers are placed at the upper or lower end of the chambers,
away from the beam pipe at a location where the radiation load is
below 50 Gy per year. The TDC electronics is installed on the
outer frames, at an even lower radiation level.

\section{Design Criteria for  the Front-End Electronics}
\label{secthree}
 
The performance of the front-end electronics influences the drift chamber
resolution, the detection efficiency and the rate of noise hits. In
the following we discuss the considerations which led to the definition
of requirements for the Outer Tracker electronics.

\subsection{Requirements on the Front-End Electronics}
\label{secthreeone}

A good position resolution of the drift chamber hits is not only needed
for a precise momentum measurement, but in a dense particle
environment it also facilitates the pattern recognition. For the Outer
Tracker it was estimated that the resolution should be about 200
$\mu$m \cite{herab}. 

In order to achieve this for the given cylindrical drift tube geometry
and the ionisation density of the chosen gas the electronics should be
able to trigger on the first cluster of electrons arriving at the
anode. The charge signal generated by the electrons at the amplifier
input depends on the gas gain.  For safety reasons, to avoid
high-voltage break-down and accelerated aging effects, it was decided
to limit the gas gain to $3\cdot 10^{4}$. With this gain an average
cluster (including the attachment loss in CF$_4$) results in a charge
of about 2\,fC after fast shaping with a charge collection efficiency
of about 20\% which is typical for drift chamber electronics. With a
threshold corresponding to a single electron cluster also the
requirement of the First Level Trigger of a high single cell
efficiency (design value $> 98\%$) can be fulfilled. The efficiency
should stay high for counting rates of up to about 2 MHz per cell
corresponding to the maximal allowed occupancy.

Processing such small signals requires low noise and low crosstalk in
the system.  The noise occupancy per channel, i.~e.~the probability
to find in a channel a noise hit in the readout window of 96~ns,
should not exceed 1\% to limit the amount of false data.
The crosstalk of a charge signal into neighbouring channels depends on
its total charge which is roughly 40 times larger than the desired
threshold charge. Therefore the analog crosstalk should not exceed
1\%.

The strongest demands on the amplifier bandwidth and the signal
shaping come from the First Level Trigger which requires a fast signal
collection within the bunch separation time of 96\,ns. With a fast signal
shaping also pile-up in following bunches will be reduced.

Because of the large number of channels a high integration density of
the amplifiers at low power consumption is necessary. The requirements
on radiation tolerance are moderate, as explained in the previous section.

For the digitization of the drift time 1\,ns bins are sufficient
because in this case the uncertainty of the time measurement adds only
about 25\,$\mu$m to the position resolution which is negligible when
quadratically added to the intended 200\,$\mu$m.  To allow for a
deadtime-less trigger and readout scheme the data have to be stored in
a 128 bunch crossings deep pipeline \cite{daqpaper}.

In the design phase of the electronics, which started in 1995, the
evaluation of available electronics led to the selection of the
amplifier-shaper-discriminator chip ASD-8 and the decision to develop
a TDC chip as the basic components of the front-end electronics.


\subsection{Overview of the system}

\begin{figure}
     \begin{center} 
\leavevmode
\epsfxsize=13cm  \epsfbox{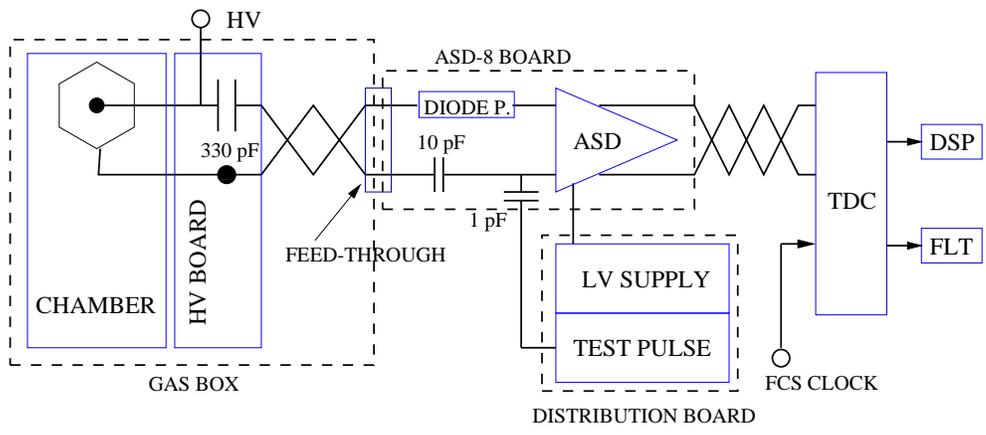} \\[1cm] %
\epsfxsize=13cm  \epsfbox{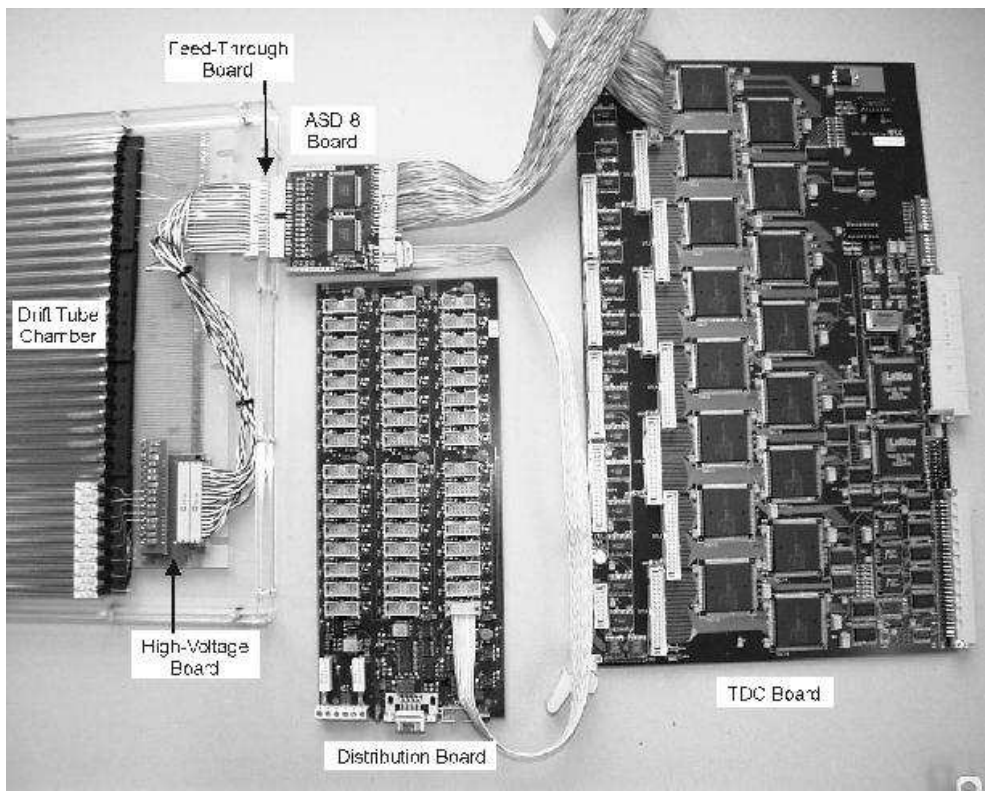}%
         \end{center} 
\caption{The front-end electronics of the Outer Tracker 
  (top: schematics for one channel, bottom: photograph of a demonstration
  assembly cabled for 16 channels).}
\label{figchain}
\end{figure}  

The front-end electronics of the Outer Tracker (fig.~\ref{figchain})
consists of
\begin{itemize}
\item[-] a high-voltage (HV) board with coupling capacitors for the
chamber signals and resistors for the current protection,
\item[-] a twisted pair cable, routing the signals to a feed-through
board in the wall of the gas box,
\item[-]a feed-through board which transfers the signal through the
  wall of the gas box and which serves as a mount for the ASD-8 board,
\item[-]an ASD-8 board,
\item[-]a low voltage distribution board to supply the ASD-8 board
with power, threshold voltages, and test pulses,
\item[-]a shielded twisted pair cable from the ASD-8 board to the TDC
board,
\item[-]a TDC board.
\end{itemize}

The HV boards, distributing the HV to each cell, are mounted on the
modules inside the counting gas-volume where the dry gas serves as
discharge protection.  The other readout electronics components are
mounted outside the gas box for better accessibility. The HV part
includes a coupling capacitor for each channel which AC-couples the
anode signals to the amplifier since the anode wires are on positive
high voltage while the cathode foils are connected to the ground. The
TDC boards are housed in crates on the outer frames of the superlayers
which carry the cables between the ASD-8 and the TDC boards.  The TDC
boards are connected via electrical cables (signal standard: LVDS =
Low Voltage Differential Signaling) to digital signal processors which
build the front-end of the DAQ system.  In addition the hit
information of 4 selected superlayers (trigger layers) is transferred
to trigger link boards for use in the First Level Trigger system. More
information on the HERA-B trigger and data acquisition system can be
found in \cite{daqpaper}.

In the following section we discuss the functionality of the listed
components. More details, including the electronic schematics of the
developed boards can be found in \cite{schematics}.

\subsection{Grounding and Shielding}
\label{sec_ground}
A concept for grounding and shielding, which is essential for the
performance and stability of the whole system, has been developed to
minimize noise, pickup, crosstalk and signal feedback.

 \begin{figure}
    \begin{center} \leavevmode
\includegraphics[clip,width=\textwidth,bb=35 105 594 491]{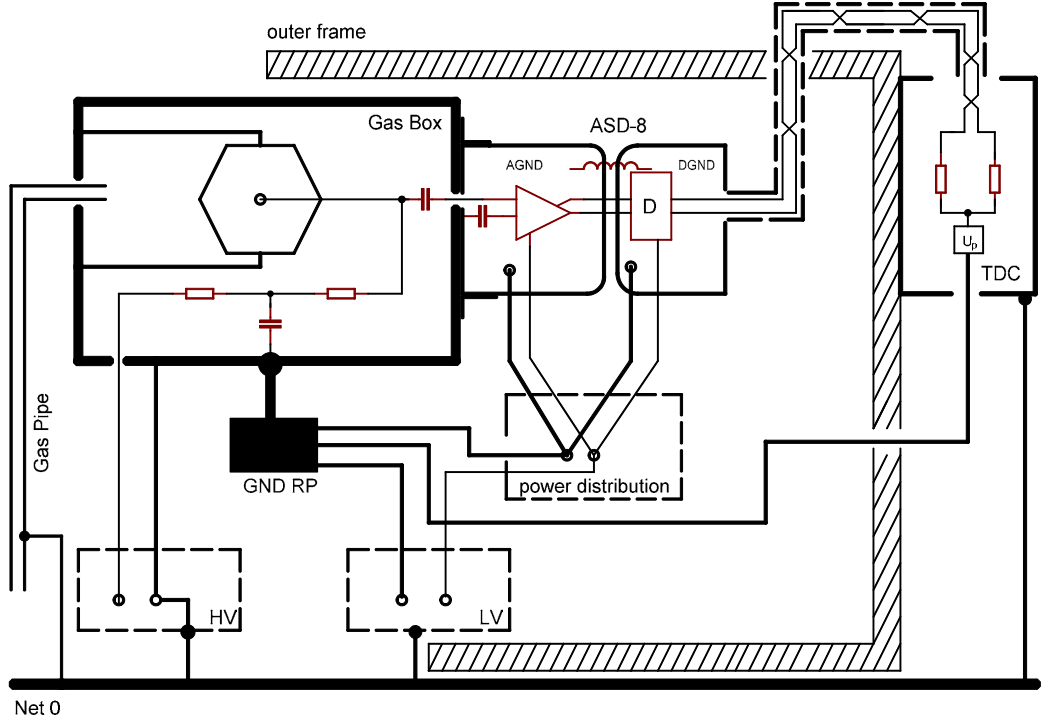}
\end{center} 
\caption{Grounding scheme of the Outer Tracker front-end electronics.}
\label{figground}
\end{figure}  
In the following we explain the scheme  referring
to fig.\,\ref{figground}. For each superlayer half  one single
``ground reference point'' (GND RP) is defined which is chosen to be
fixed to the potential of the gas box. The ground
potentials of all components have to be connected to GND RP
(star-like).  There is only one connection to the reference potential
of the mains (Net0) via the HV power supply which, for safety reasons,
had to be connected to the gas box. All other lines coming from Net0
connect only to otherwise electrically insulated parts, like the
enclosures of electronics (LV, TDC), the outer frame, and the gas pipe.

The most critical points are the grounding of the analog and digital
sections on the ASD-8 board, and the signal connections from the
chamber to the ASD-8 board and from the ASD-8 board to the TDC. On the
ASD-8 board, the digital ground (DGND) and the analog ground (AGND)
are separated (see section \ref{secasd_board}).  A good connection of
the AGND to the cathodes is mediated by the gas box enclosing each
chamber. From the inside, the cathodes are connected to the box, and
from the outside Cu-Be brackets, carrying the ASD-8 boards, connect to
the AGND of the board. Being designed as a Faraday cage the gas box
serves also as a RF shield. It is insulated from the outer frame and
from the gas pipe (using insulating pipe connections).

Shielding of the signal cables going to the TDC turned out to be
absolutely necessary to minimize the feedback of the digital output
signals to the amplifier inputs. The shielding is connected to DGND on
the ASD-8 board without any direct ground connection from the ASD-8
board to the TDC. The ground potential on the TDC board is connected
to the ground reference point GND RP.

The final tuning of the grounding and shielding of the complete system during
commissioning is described in  section \ref{secinstall}.


\section{High Voltage System}

\subsection{High Voltage and Signal Routing in the Gas Box}
\label{sec_hv_routing}

\begin{figure}
     \begin{center}
\leavevmode
\epsfxsize=9cm  \epsfbox{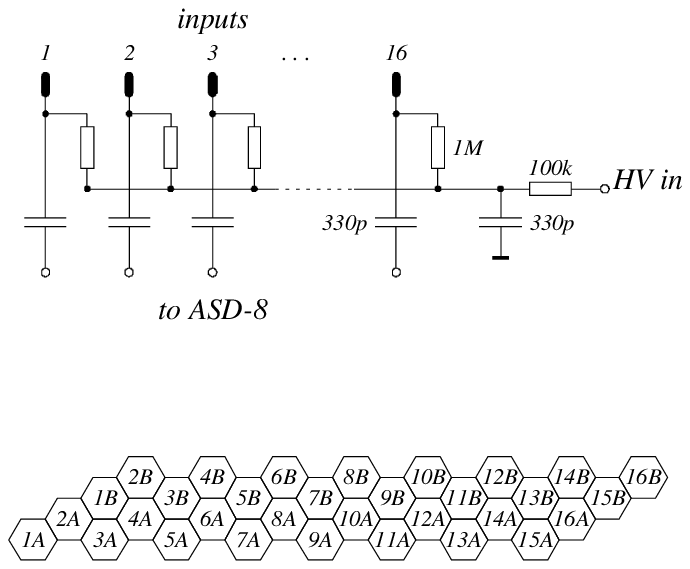} %
         \end{center}
\caption{ High voltage distribution board: schematics (top)
  and board-to-chamber mapping (bottom).}
\label{fighvboards}
\vspace{1cm}
%
%
     \begin{center}
\leavevmode
\epsfxsize=9cm  \epsfbox{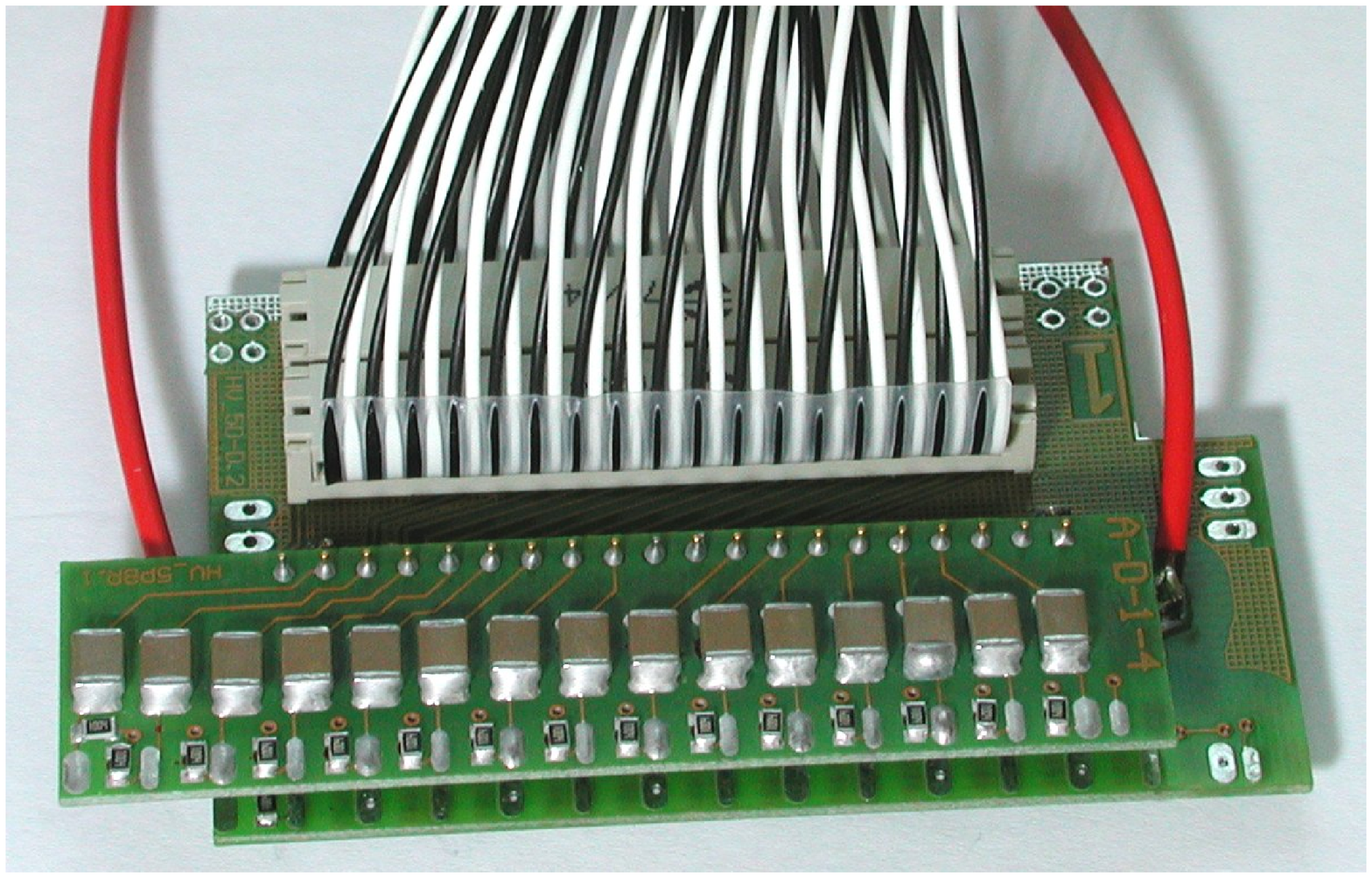} %
         \end{center}
\caption{ High voltage distribution board:
  a photograph of the connected main and piggy-back boards for the
  5~mm chambers.}
\label{fighvboardp}
\end{figure}

The high voltage is supplied to the anode wires via HV boards, which
are mounted on the module base plates \cite{otr_det}. The schematics
of the boards is shown in fig.~\ref{fighvboards}.  Besides supplying
high voltage to individual anode wires via a 1 M$\Omega$ protection
resistor, the board also leads the signals from the anode wires through
coupling capacitors (330\,pF, 4\,kV)\footnote{ceramic chip capacitors
330\,pF, 4\,kV, X7R dielectric, Johanson Dielectrics (art.nr.~402 S43
W 331 KV4)} to the cable connectors, leading to the ASD-8 boards. The
high voltage enters the board through an RC filter (100 k$\Omega$,
330\,pF).

To accommodate the HV board to the chamber structure
(fig.~\ref{fighvboards}), a combination of two boards, a main
board with a piggy-back board on top of it, is used
(fig.~\ref{fighvboardp}). The main board supplies the channels 1A to
16A, the piggy-back board the remaining channels 1B to 16B.  For the
chambers which are not used in the First Level Trigger the signal
routing is such that the channels on the main board end up in one
cable and the channels from the piggy-back board in another. For
trigger chambers the cabling is such that back-to-back drift cells nB
and (n+2)A end up in neighbouring channels of the output cable. This
routing facilitates performing a logical OR of hits from these cells
in order to increase the trigger efficiency.

The HV boards are double-sided printed circuit boards with surface mounted
components (SMD). For the 5\,mm cells the matching to the wire pitch
limits the board width to 67\,mm requiring a tight spacing between the
capacitors. In order to guarantee the high voltage proofness, two of
the 17 HV capacitors are placed on the back side of the board (because
of a different soldering method these two capacitors caused major
problems in the first running period, see section \ref{secinstall}).

The signal cables (16 twisted pairs, lengths between 25 and 50 cm) from the HV
boards are plugged on the inside of the gas-box to the feed-through
boards which hold on their outside the ASD-8 boards.  The feed-through
boards provide also the feed-through for the high voltage and a
possibility to individually disable problematic HV boards.

%

\subsection{The High Voltage Distribution and Protection System}

The Outer Tracker High Voltage System has a cascaded distribution
scheme which optimizes the number of dead channels in case of high
voltage problems due to unstable or broken wires with respect to the
cost of the system.  It provides positive voltage for the 112\,674
anode wires, nominally 1700\,V for 5 mm drift cells and 1950\,V for
10\,mm cells.

The HV grouping follows the structure of the Outer Tracker which is
described in \cite{otr_det}: The largest individual detector units are
the superlayer halves, being composed of stereo layers (6 for PC and 3
for TC) which are then subdivided into sectors (12 per PC, 10 per TC
stereo layer).

The HV supply system\footnote{CAEN SY527 Multichannel HV System: One
crate with 10 HV Supply Boards A734P, each board with 16 HV channels
(max. 3\,kV/3\,mA), connected to the Slow Control system via V288
HS-CAENET-VME Interface.} has 160 individually controllable HV
sources, each of which feeds up to six sectors with
up to 1500 wires in total. Such a
 HV source is the smallest unit which can be monitored and controlled
 by the Slow Control software.  However, failing single sectors can be
 disconnected individually on a patch board in the electronics
 trailer. Beyond that also the groups of 16 wires belonging to an HV
 board (fig.\,\ref{fighvboards}) can be switched off at the
 feed-through board on the gas box frame. This action requires an
 access to the detector which is regularly scheduled once per
 month. In this way the number of disabled channels per faulty wire
 can be limited to 16.

For each of the 160 HV sources the current is continuously monitored
and in case of over-currents a protection scheme first reduces the
voltage and eventually switches it off.  An interlock makes sure that
the HV can only be switched on, if the gas system works properly. To
provide a constant gas gain the HV is adjusted automatically within
defined limits by a gas monitor system. A description of the control
system is given in \cite{otr_perf}.

\section{The ASD-8 Electronics}

\subsection{The ASD-8 Chip}

The amplifier-shaper-discriminator chip ASD-8 \cite{asdpub}, developed
by the University of Pennsylvania for drift chamber applications in a
high rate environment (originally for SSC), is used in different
detector systems of the HERA-B experiment (Outer Tracker, RICH, Muon
System, High-p$_T$ Chambers) with a total of about 200\,000 channels.

\begin{table}
\begin{center}
\caption{Characteristics of the ASD-8 chip.}
\vspace{1mm}
\label{tab01}
\begin{tabular}{|l|l|}
\hline
integration density    & 8 ch. on $2.7\times 4.3$\,mm$^2$ die \rule{0.0mm}{0mm} \\
power consumption & 0.3 W\,/\,chip (incl.\,output)\\
signal shaping time & about 10\,ns \\
tail cancellation & $t_0 \approx 1.5$\,ns\\ 
double pulse resol. & 25\,ns \\
intrinsic noise & (900 + 70/pF) electrons\\
on-chip crosstalk & analog: $\sim$ 0.1\% \\
threshold & $\sim 2\,$fC  \\
baseline shift & 0.5 - 1.0 fC/MHz \\
\hline
\end{tabular}
\end{center}
\end{table}

The 8-channel chip is designed in
\begin{figure}
    \begin{center} \leavevmode%
\epsfxsize=10cm  \epsfbox{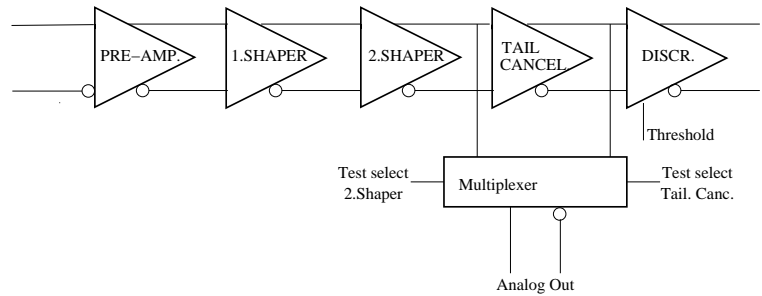}
\end{center} 
\caption{Principal functions of the ASD-8 chip. The analog outputs 
  are only used for test purposes and can be selected by an on-chip
  multiplexer.}
\label{figasdprinzip}
\end{figure}  
a bipolar technology. The  process used by the producer Maxim combines high
speed with a low noise level, 
and relatively
low power consumption (Table \ref{tab01}).

A block diagram of the  ASD-8 chip is shown in fig.\,\ref{figasdprinzip}.
The input stage is a preamplifier with a sensitivity of 2.5\,mV/fC, a
bandwidth of 100 MHz and an input impedance of 115\,$\Omega$. The input
is differential and symmetric for positive and negative pulses.

The two-stage shaper with tail cancellation yields a double pulse
resolution of 25\,ns. The tail cancellation compensates the ion tail of
the drift chamber pulses.  Analog outputs are provided for 3 channels
per chip, selectable after the second shaper or after the tail
cancellation.

The discriminator is a two-stage differential amplifier with positive
feedback. The threshold is voltage programmable for each channel. The
external voltage scales approximately like 250 mV\,/\,fC up to about
1.4 V where the threshold saturates. The baseline shift given in Table
\ref{tab01} is not negligible at the highest occupancies and has to be
compensated by a corresponding threshold shift.  The differential,
open collector output is current programmable to adjust the signal
swing.

\subsection{The ASD-8 Board}
\label{secasd_board}

\paragraph*{Design considerations:}
The analog inputs of the ASD-8 chips have a very high sensitivity
which makes them susceptible to noise and RF
pickup. Because of the combination of analog inputs and digital
outputs on the chip a particular worry is the feedback from
\begin{table}
\begin{center}
\caption{Technical data of the ASD-8 board}
\vspace{1mm}
\label{tab03}
\begin{tabular}{|l|l|}
\hline
channels & 16 (2 chips)\\
dimensions & 4 layers,  67 $\times$ 56 mm$^2$  \\
pickup suppression &  all supply voltages RC filtered \\
spark protection & diode protection $\geq$3 kV \\
crosstalk (analog) & $<0.5\%$\\
grounding & analog - digital separated \\
gain uniformity &  $\pm 15\%$ per board \\
output & 2\,mA into 62\,$\Omega$ \\
voltage supply &  $+3$ V, 100 mA \\
        &  $-3$ V, 100 mA \\
power & 600 mW  \\
\hline
\end{tabular}
\end{center}
\end{table}
the output to the input. In the design of the printed circuit boards
carrying the ASD-8 chips special care was taken for a good grounding
scheme, decoupling of the analog and digital parts, noise rejection
from power sources, and crosstalk separation of different channels in a
densely packed environment.

Since only one board type for both the 5\,mm and 10\,mm cells should
be produced, the geometrical constraints are defined by the dimensions
of the 5\,mm cells. The width of the boards was adjusted to the cell
pitch; a compact assembly of connectors and electronic components was
achieved using SMD technology.

\begin{figure}
\begin{center} \leavevmode
\epsfxsize=8cm  \epsfbox{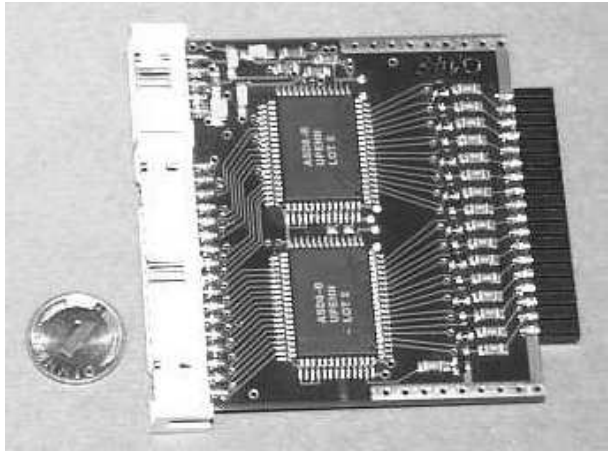}
\end{center} 
\caption{Picture of the ASD-8 board of the  HERA-B Outer Tracker.}
\label{figasdphoto}
\end{figure}

\begin{figure}
\begin{center} \leavevmode
\epsfig{file=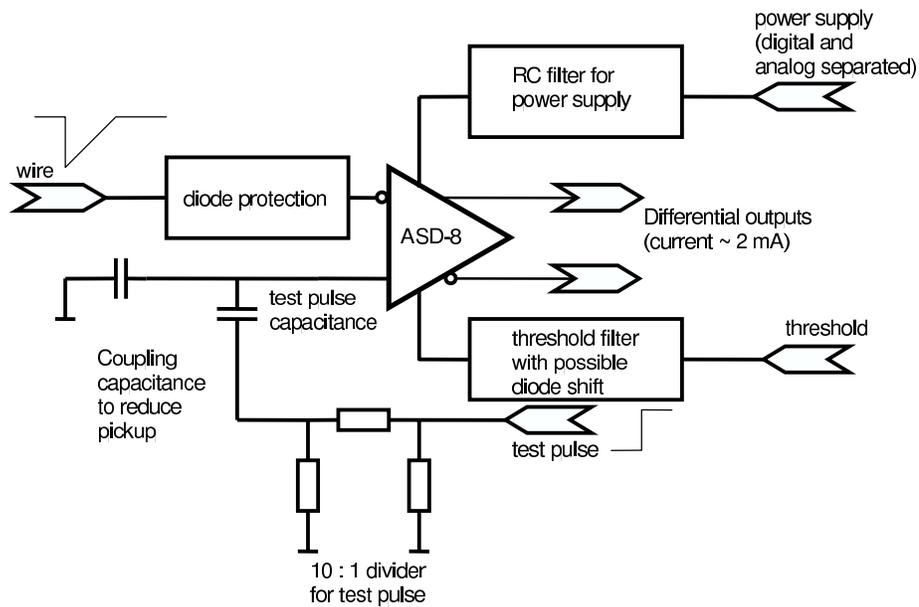,width=8cm,angle=-90,}
\end{center} 
\caption{Block diagram of the ASD-8 board.}
\label{figschematic}
\end{figure}

%
\paragraph*{Board layout:} Two ASD-8 chips with 8 channels each are
mounted on a multilayer board (Table \ref{tab03},
fig.\,\ref{figasdphoto}).  A block diagram of the board is shown in
fig.\,\ref{figschematic}, more details can be found in \cite{schematics}.

The chip has a differential input while the nature of the chamber
signals is not differential. With the anode signal fed into the
negative input different options for the positive input were
tested. The best common mode rejection
was obtained by connecting the positive input via 10\,pF to the
chamber ground, i.~e.\ the cathode (fig.\,\ref{figchain}). This
quasi-differential use of the inputs was clearly superior to the
option to leave the positive input open.

\begin{figure}
\begin{center} \leavevmode
\epsfxsize=12cm  \epsfbox{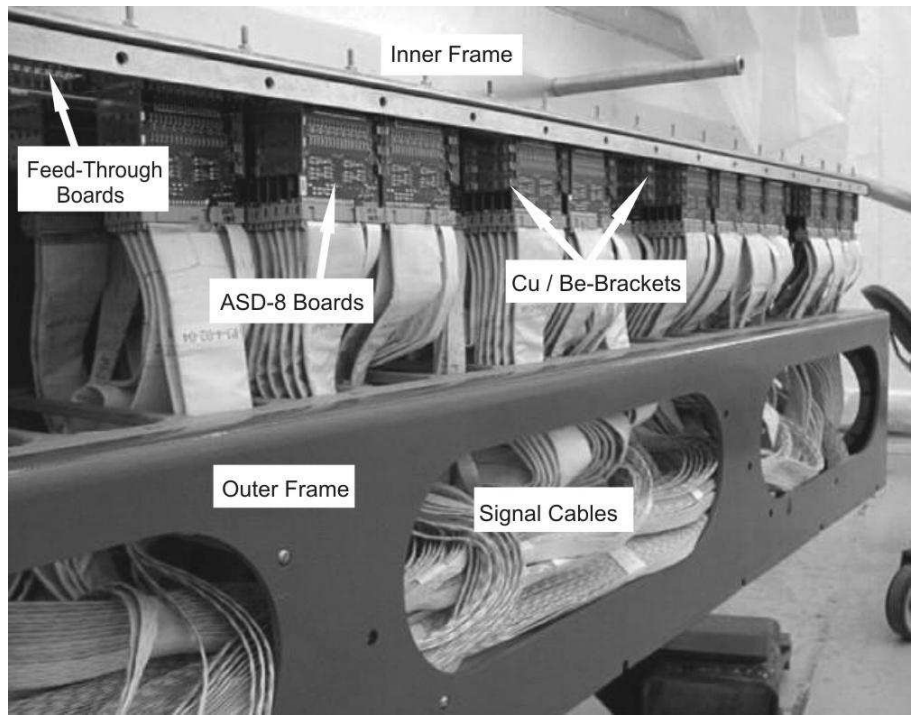}
\end{center} 
\caption{Photograph of the front-end electronics on an  Outer
Tracker chamber: Shown is the lower part of a chamber with the ASD-8
cards plugged onto the feed-throughs at the gas box, the cables
connecting the ASD-8 outputs with the TDCs and the cable frame routing
the cables to the TDC crates (on the vertical frame part to the left,
not visible here).}
\label{figfeephoto}
\end{figure}

All supply voltages ($\pm 3$\,V) are separated for the  analog, digital
and output drive circuits and have RC filters at the input. The ground
plane is separated for the analog and digital part of the chip. Both
grounds are connected via an inductance of $10\,\mu$H. The analog ground
is extended to rails along the sides of the board which slide into the
holding brackets on the chambers made of Cu-Be springs. Thus the
brackets together with the ground planes provide a shielding mesh
between the boards
in the densely packed front-end electronics on the chambers (see
fig.\,\ref{figfeephoto}).
Since it was found that the ASD-8 inputs survive voltage spikes only
up to about 300\,V, a diode protection circuit was added to protect
against high voltage breakdowns in the drift cells. The protection
circuit consists of an input resistor of $50\,\Omega$ and two diodes
connected in parallel with opposite polarity (type BAV79) shortening
the ASD-8 input against a 0.7\,V level \cite{schematics}.  Tests have
shown that in this way the input transistors could be protected
against discharges of 3000\,V fed into the input via a 1\,nF
capacitor.

Only one voltage level for the thresholds is provided per board.  The
two chips on each board were therefore selected in order to be in the
same gain and noise quality classes (defined in
section \ref{secchiptests}).
The maximum difference of the reference thresholds $U_{ref}$ of two
channels of a single board, as defined below, should be less than
300\,mV. In some cases Schottky diodes (forward voltage 200 or
380\,mV) were used to shift the
thresholds of chips or of single channels to avoid more categories or
to reduce the number of rejects.

\begin{figure}
     \begin{center} 
\begin{tabular}{c}
 \epsfxsize=8cm  \leavevmode\epsfbox{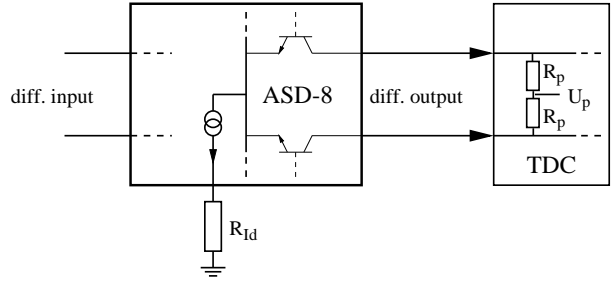}
\end{tabular}
\end{center} 
\caption{ Schematics of the open
collector, differential output of the ASD-8; the current is adjusted
by the resistor $R_{Id}$. The output current together with the pull-up
resistors $R_P = 62\,\Omega$ and the voltage $U_P = 1.25$\,V
determines the signal swing and the reference level of the signals.}
\label{figasdlevel}
\end{figure}

The current of the open collector, differential output of the ASD-8 is
adjustable. It was chosen to yield a swing of 120\,mV for the given
pull-up resistors on the TDC boards (fig.~\ref{figasdlevel} and
Sec.~\ref{tdc_circuit}). An offset of 1.25\,V is added by the
receivers on the TDC board. The analog outputs of the ASD-8 chips are
not used.

Test pulses are capacitively coupled (1\,pF) into the positive input
via a 1:10 voltage divider to reduce noise pickup via the test pulse
system.  The pulses fire all channels on a board at the same time.

\subsection{Low Voltage Distribution Board}
\label{secpdb}
The low voltage distribution board supplies 6 groups of 8 ASD-8 boards
each with power, thresholds and test pulses
(fig.\,\ref{figchain}). These boards are controlled by a SLIO
processor (Serial Linked I/O; Philips P82C150) and are connected to
the HERA-B Slow Control System via a CAN (Controller-Area-Network)
bus.

On each board, two different thresholds can be set using 10\,bit
DACs. For each of the 6 groups one of these thresholds has to be
preselected by setting a jumper. The board has an RS422 input for test
pulses which can be enabled for each group separately by the CAN
controller. The pulses have a length of about 30\,ns; the amplitude
can be adjusted by a potentiometer to a common value for each
group. An internal ADC of the SLIO measures the real thresholds,
supply voltages and currents using a multiplexer.  The analog and
digital supply voltages for each ASD-8 board are filtered by RC
circuits.

\subsection{Test and selection of ASD-8 chips and boards}

\subsubsection{Series Chip Tests and Selection}
\label{secchiptests}

\paragraph*{Definition of chip quality: }
HERA-B ordered a total of 90 wafers, each with about 335 chips. The
yield per wafer was on average 77\%. Each chip was tested according to
a scheme which evaluated the general functioning, the threshold
behaviour, and the noise level.

For each channel a threshold reference voltage, $U_{ref}$, was defined
as the threshold at which a standard 4\,fC test pulse is recorded with
50\% efficiency.  The noise behaviour of a chip was characterized by
the threshold $U_{noise}$ for which the noise rate exceeded 2~kHz. The
difference $U_{ref} - U_{noise}$ is a measure for the signal-to-noise
distance and thus of the quality of a channel. The parameters of the
test (4\,fC, 2\,kHz) were chosen to yield a high test sensitivity in
the relevant range, 
and to yield stable and reproducible results.

\begin{figure}
    \begin{center} \leavevmode
\begin{tabular}{cc}
\epsfxsize=6.5cm  \epsfbox{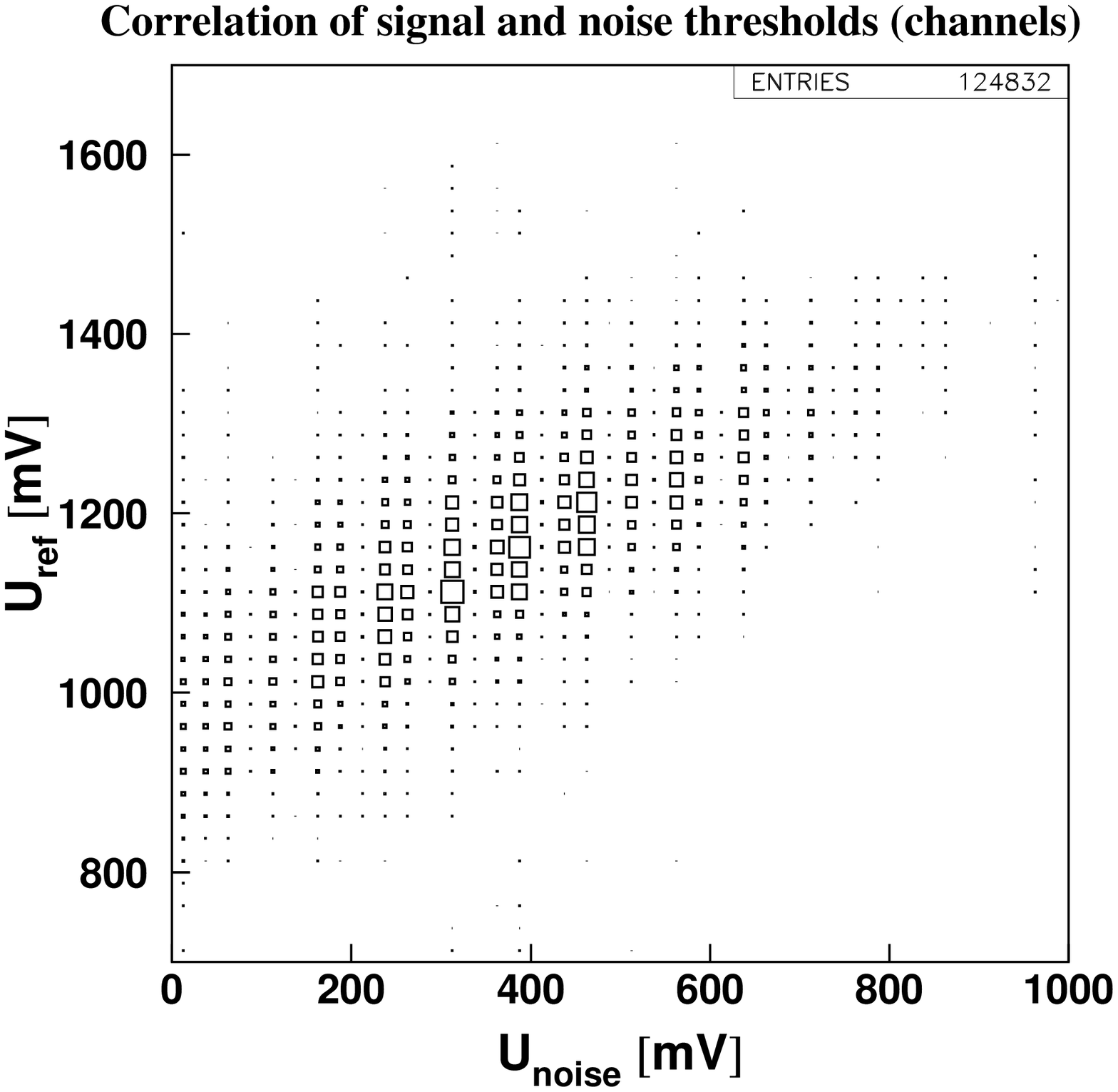} &
\epsfxsize=6.5cm  \epsfbox{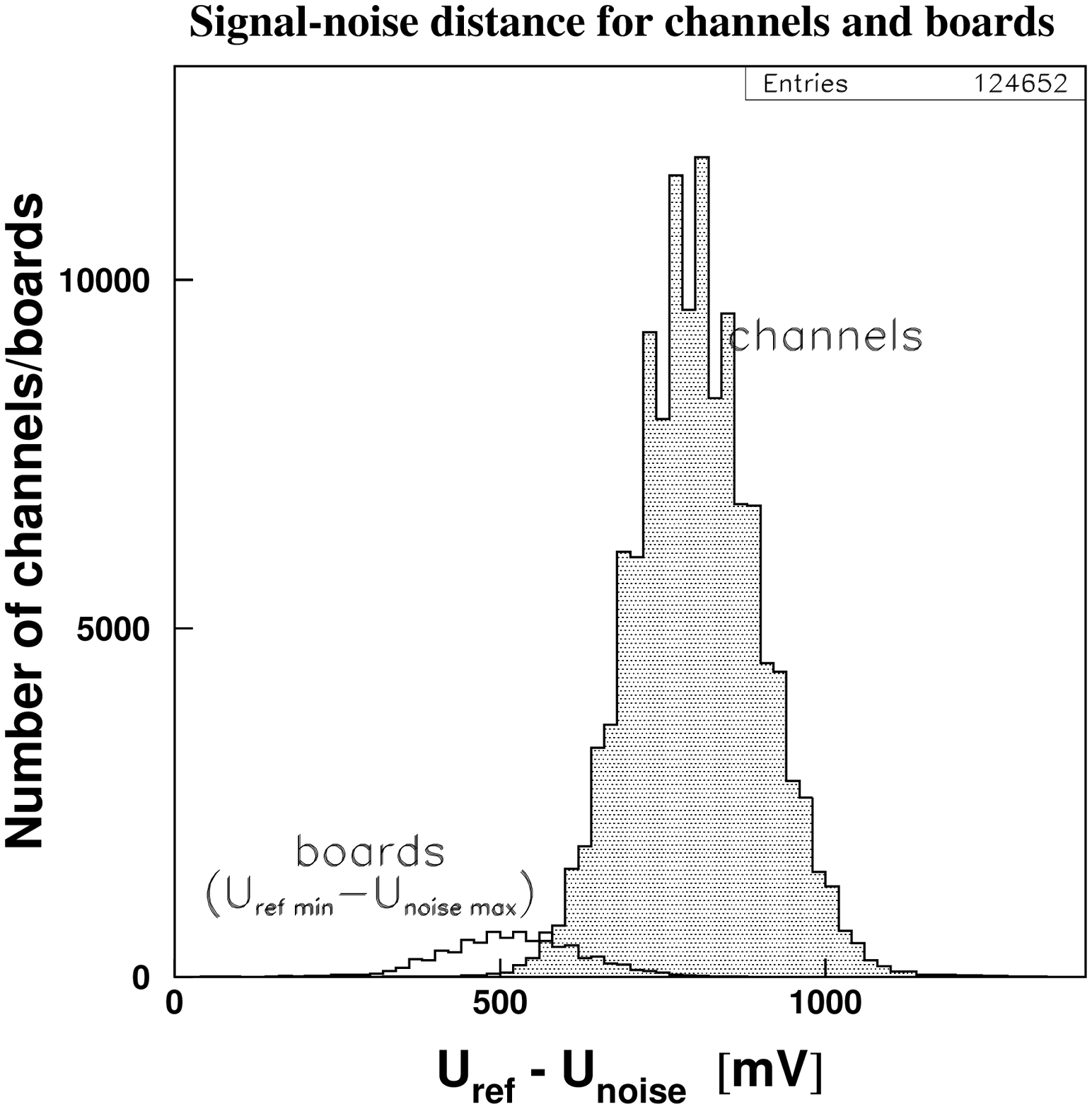}%
\end{tabular}
\end{center} 
\caption{Left: Threshold $U_{ref}$ plotted against the noise
threshold $U_{noise}$ for each channel of the tested chips. Right:
Distribution of the signal-noise difference $U_{ref}- U_{noise}$ for
each channel of the tested chips and for the boards where the
signal-noise difference is defined by the channels with minimal
$U_{ref}$ and maximal $U_{noise}$. }
\label{figcorr}
%
\vspace{5mm}
\begin{center} \leavevmode

\epsfxsize=6cm  \epsfbox{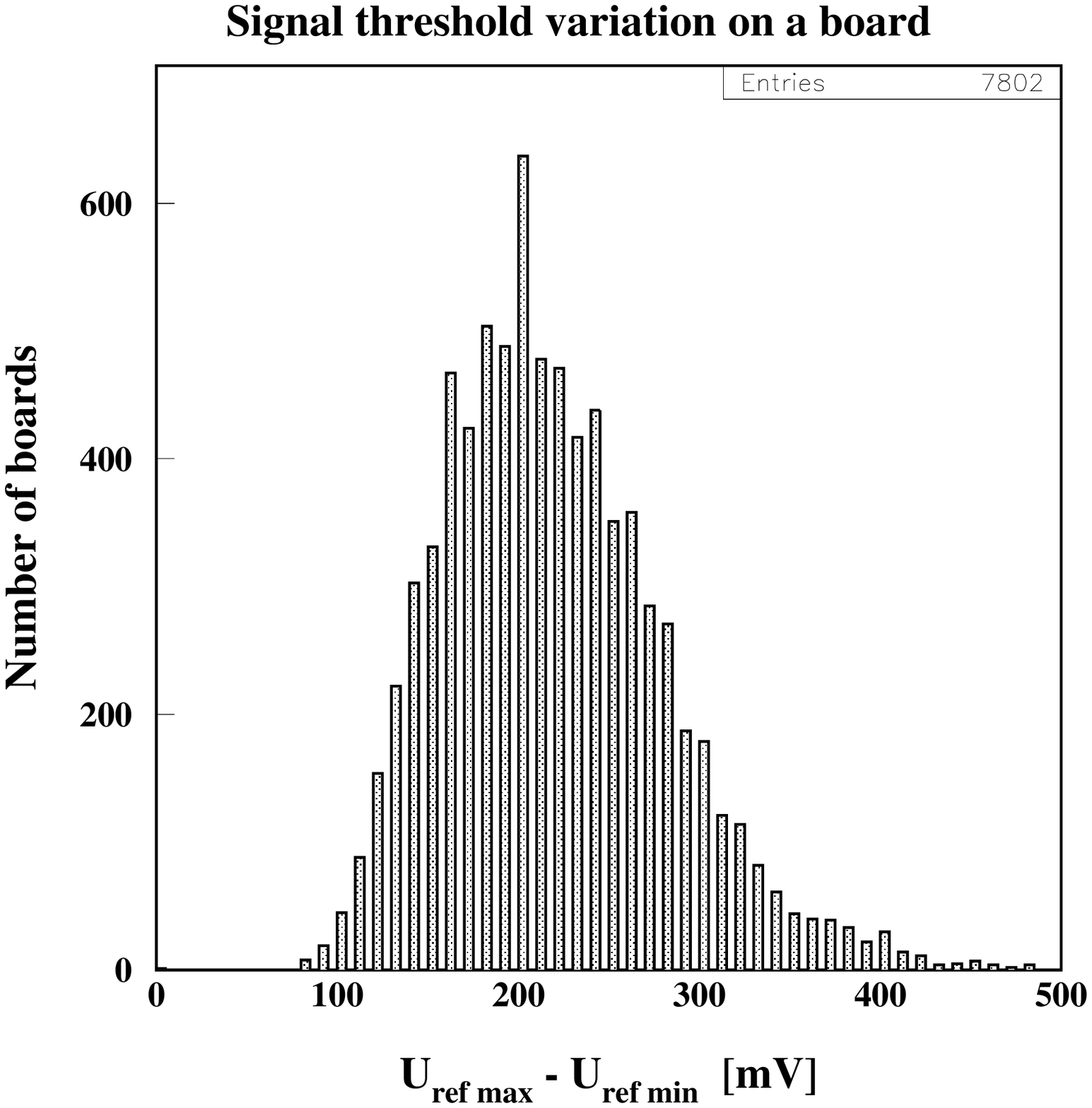}%
\end{center} 
\caption{Threshold uniformity on the ASD-8 boards: distribution of the
  difference between maximal and minimal reference threshold $U_{ref}$
  on a board ($\sim 250\,$mV/fC). 
Differences above 300\,mV have been reduced by a diode circuit (see text).
}
\label{figdiffuthr}
\end{figure}

\paragraph*{Results of the chip test: }
The chip tests revealed an appreciable range for the $U_{ref}$
threshold (fig.\,\ref{figcorr}). However,
within one chip the variation of thresholds was found to be mostly much
less than the $\pm 30\%$ specified in the purchase order.
The chip-to-chip variations prohibit the definition of a common
threshold for the whole system  keeping at the same time the
overall efficiency high.
 On the other hand, to keep the front-end electronics simple and
compact, individual threshold settings for each channel have to be
avoided. As a compromise 4 categories of threshold ranges were defined:
$$
U_{ref} = 850 \ldots 950,\, 950 \ldots 1050,\, 1050 \ldots 1150,\, >
1150\,{\rm mV}.
$$
To account for the signal-to-noise variation for a given $U_{ref}$
(fig.~\ref{figcorr}) in each of the 4 threshold categories, 3 noise
categories corresponding to noise distances were defined:
$$
 \overline{U_{ref}}-U_{noise}  = 
350 \ldots 450,\, 450  \ldots 550,  \, > 550\,{\rm mV},
$$
where $ \overline{U_{ref}}$ are the central values of the  threshold
categories ($\overline{U_{ref}}$ = 900, 1000, 1100, 1200\,mV).  A chip
was assigned to one of the 12 categories according to its minimal
$U_{ref}$ and maximal $U_{noise}$ values. The assignment was used to
mount two similar chips on a board and to combine similar boards to
groups which are supplied with a common threshold.

\subsubsection{Board Tests}

The 11000 produced boards had to undergo quality tests and were then
sorted according to 12 categories as was done for the single chips: 4
categories according to the threshold reference voltage, $U_{ref}$,
and 3 categories of signal-to-noise distance $
\overline{U_{ref}}-U_{noise}$ (see fig.\,\ref{figcorr} right). A board
enters into a category according to the minimal $U_{ref}$
and maximal $U_{noise}$ of the channels.

Details about the distribution of the boards in different categories
are given in \cite{lhce}. For each board the two chips belong to the
same category. The remaining variations of $U_{ref}$ within the 16
channels can be seen in fig.\,\ref{figdiffuthr} which shows the
difference between the maximal and minimal $U_{ref}$ on the boards.
Differences larger than 300 mV have been decreased by Schottky diodes
as described in section \ref{secasd_board}. With this
procedure the threshold uniformity on the boards is about $\pm 15\%$.
The variations between different boards are accounted for by the
threshold settings on the distribution boards (section \ref{secpdb}).

The board quality is mainly determined by the noise
category, more than 80\% are in the upper two categories with
$\overline{U_{ref}}-U_{noise}  > 450 \,{\rm mV}$ for both chips.

\renewcommand{\textfraction}{.3}
\renewcommand{\floatpagefraction}{.7}

\section{The Time Measurement System}

\subsection{Introduction}

The HERA--B TDC system has been developed for the readout of the Outer
Tracker, the RICH, the Muon System and the High-p$_T$ Chambers. 
Except for the Outer Tracker the other systems use the TDC chip only as
a hit register.  The system is highly integrated at low power
 consumption and reasonable costs (Table \ref{tab02}).

\begin{table}
\begin{center}
\caption{Characteristics of the TDC chip.}
\vspace{1mm}
\label{tab02}
\begin{tabular}{|l|l|}
\hline
integration density    & 8 ch. on $9.4\times 9.4$\,mm$^2$ die \rule{0.0mm}{0mm} \\
power consumption & 0.06 W\,/\,chip\\
operating voltage & 5 V\\
time bin  & about 0.5\,ns \\
time resolution & about 0.2\,ns\\
\hline
\end{tabular}
\end{center}

\end{table}
\begin{figure}
  \begin{center}
\epsfig{file=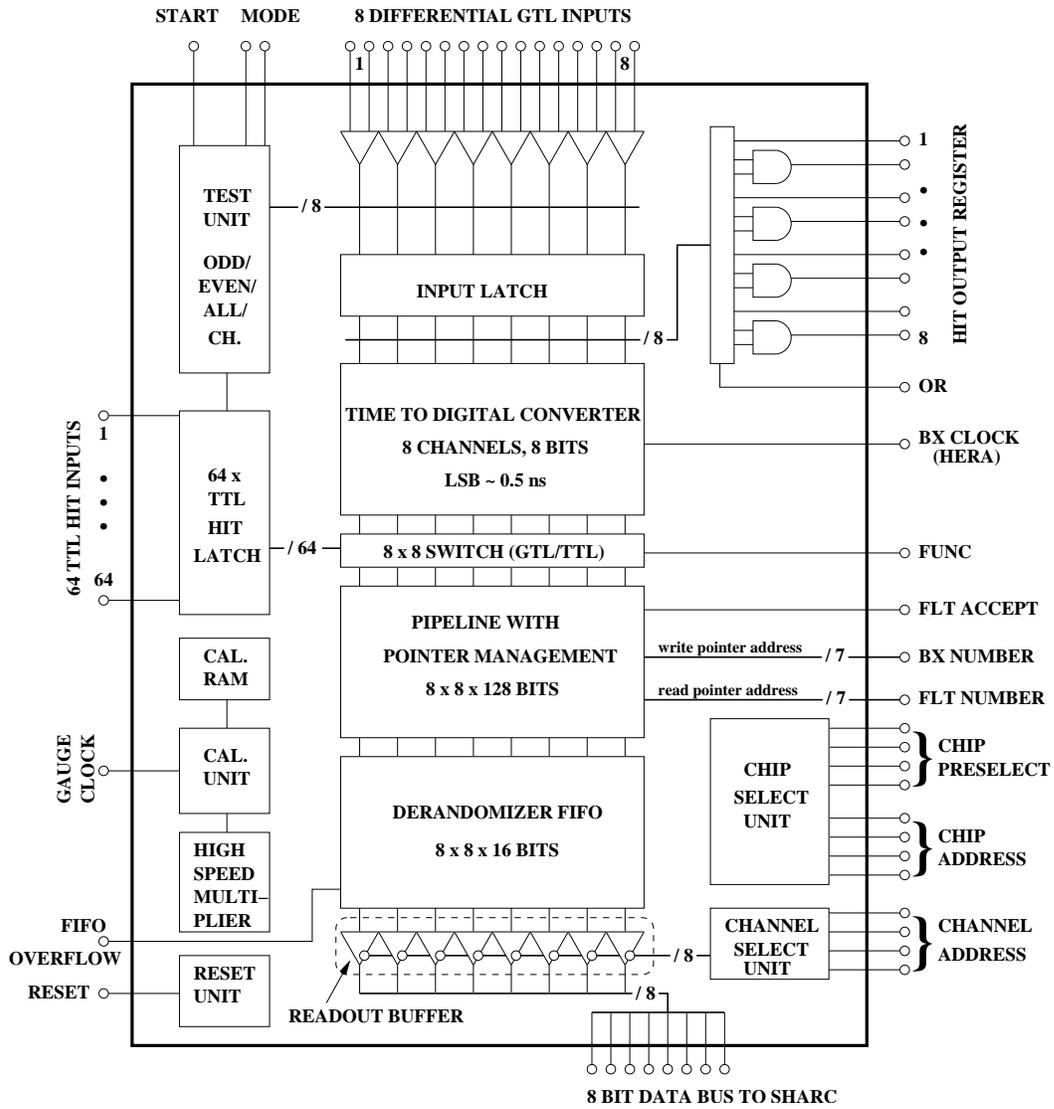,width=\textwidth}
    \caption{\label{chip} Block diagram  of the TDC chip.}
  \end{center}
\end{figure}
 
The TDC was designed to digitize the time in 0.5\,ns bins with an 8
bit output.  After digitization the data of each channel is stored in
a 128 cells deep digital pipeline which is read out by a digital signal
processor (SHARC processor ADSP-21060 from Analog Devices).  With the
integrated channel buffering the requirement of a dead-time free
readout after the First Level Trigger decision is fulfilled.

The chips are mounted on boards with 16 chips each, corresponding to 128
TDC channels per board.  The crates housing the TDC boards are
mounted on the cable frame of the detector superlayers, except for the
magnet chambers for which the TDC crates are placed outside the
magnet.
The maximal cable length between the ASD-8 boards and the TDC crates
is 5 to 10 m, depending on the superlayer.

\subsection{\label{tdc_circuit}The TDC Chip}

\subsubsection{Overview of the Layout}

Figure \ref{chip} shows the structure of the TDC chip which was
designed by the company MSC\footnote{MSC, Industriestr.\,16, D-76297
  Stutensee, Germany} and produced by the company NEC (Japan). The chip was
developed as an ASIC (Application Specific Integrated Circuit) in
0.8~$\mu$m CMOS technology
using VLSI (Very Large Scale Integration) design techniques (process
provider: NEC). The 9.4~$\times$~9.4~mm$^2$ die is housed in a 160
pin-package.  The TDC chip includes the following features:

\begin{itemize} 
\item[-] 8 differential inputs for the ASD-8 outputs,
\item[-] 8 hit output registers (with optional OR for neighbouring
  channels) for the {F}irst {L}evel {T}rigger,
\item[-] 8 x 8 bit time-to-digital converter,
\item[-] calibration unit (with high speed multiplier),
\item[-] buffer memory pipeline,
\item[-] derandomizer FIFO memory,
\item[-] readout buffer with addressing unit,
\item[-] chip and channel select units,
\item[-] test device for all input signals,
\item[-] hit input register for 64 channels.
\end{itemize}

\subsubsection{Input Circuit}

The differential input signals from the ASD-8 chips are received by a
{GTL-I/O}--Interface (GTL = Gunning Transceiver Logic). 
The GTL standard, characterized as a low-level, high-speed,
noise-immune digital logic, requires the differential swing to be
$\ge$50~mV, in the Outer Tracker system it was set to 120~mV. The
levels are defined by the circuit on the TDC board as shown in
fig.~\ref{figasdlevel}. All other input/output signal levels of the
chip are TTL compatible to simplify interfacing.

After passing the GTL-I/O--Interface the drift chamber signals are
latched with the BX clock. Thus the hit information is available in
the hit output register in the next BX clock cycle and
can be used by the First Level Trigger. If required, the hit
information of two neighbouring channels can be merged by an
OR--function. 

\subsubsection{TDC Circuit and Internal Calibration}
The heart of the TDC chip is an 8 bit {TDC} circuit which converts
time differences into digitized values using a delay line chain made
of logic gates. The method\footnote{German patent nr.~41\,11\,350}
\cite{tdc_meas} is fully digital with the major advantage that no
conversion time is required and the measurements can be made nearly
without deadtime (section \ref{tdcperform}).
The time measurement bin of the circuit follows from the basic gate
delay and was measured to be about 0.48\,ns, very close to the
targeted 0.5\,ns.

Each chip has nine channels, eight for time measurements and one for
calibration (not shown in fig.\,\ref{chip}). At the startup of the
system (after a Reset) all nine channels are calibrated assuming a
linear relation between the measured time interval and the TDC output
(see section \ref{tdcperform}). For each channel, slope and offset of a
straight line are determined by measuring time intervals of 100\,ns
and 200\,ns, both derived from a 10 MHz gauge clock, with a 10\,bit
resolution. 
The straight line parameters include a mapping of a 100\,ns interval
onto the 256 output counts of the TDC.  The two parameters for each
channel are written into the memory of the calibration unit.

To compensate temperature and voltage variations the chip features a
self-calibration procedure.  During data taking, every 0.8 seconds the
slope parameter of the calibration channel is re-measured in the same
way as for the initial calibration and a correction factor is
calculated which is also written into the memory of the calibration
unit.  The time for the other 8 channels on the chip is multiplied by
this
factor. The mapping of the 100\,ns interval onto 8 bits including the
multiplication by the correction factor is done during the transfer
from the pipeline to the derandomizer by the High Speed Multiplier
unit.

The time is measured in common stop mode. The pulse from the GTL input
or the test device yields the START while the STOP is derived from the
external HERA BX clock which is synchronized with the
bunch crossing signal (BX). Because of the mapping of  100~ns
onto the 8 bits (256 counts) 
the least significant bit (LSB) of the time measurement (1 TDC count)
corresponds to 0.39~ns in HERA-B.  Note that the bins for the time
measurement remain fixed at about 0.48\,ns which determines the time
resolution (section \ref{tdcperform}).

\subsubsection{Buffer Circuits}
The {pipeline} of each channel is a ring-buffer memory with 8 bits per
cell
and a depth of 128 events. The pipeline depth corresponds to a time
interval of about 12~$\mu$s available for the First Level Trigger
decision.  Writing and reading the data is controlled by two different
pointers using the BX number as address.  On an Accept signal from the
First Level Trigger (FLT Accept) the {F}ast {C}ontrol {S}ystem (FCS)
generates the read pointer (FLT Number) and pushes the event data to
the derandomizer FIFO. The design mean trigger rate is 50~kHz although
the TDC system is capable to run at more than 100~kHz.
The {derandomizer FIFO} with a depth of 16 events serves for each TDC
channel as buffer memory for peak rates. The events in the FIFO are
read out into the {S}econd {L}evel {B}uffer (system of SHARC processors).

The readout is organized by the {Chip Select Unit} which addresses the
TDC chip and the {Channel Select Unit} which addresses the buffer of
each channel.  If the readout cannot follow the trigger rate the FIFO
is filled up and the next event could be lost. In this case a FIFO
Overflow bit is set which is used in HERA-B to stop the data
acquisition until the buffer is available again.

\subsubsection{Control and Test Functions}

The chip can be tested using an internal pattern generator which is
controlled by a 3~bit input for setting the test functions. Two bits
control the mode (all, even, odd channels on) and one bit starts the
measurement of a preselected time interval provided by the FCS system.
In this way the proper functioning of the chip can be tested.

Instead of the use as 8 channel TDC, the chip can also be used as a
hit register with 64 channels.
The hit register mode is set by a function bit (FUNC) which enables the 64
hit inputs and switches off the TDC circuits. The hit inputs are
grouped by 8, so that the same data format (8 times 8~bits) can be
used.  The buffer management is the same as for the TDC application.

\subsection{The TDC board}

\begin{figure}
  \begin{center}
    \includegraphics[scale=0.5,clip,bb=0 0 590 437]{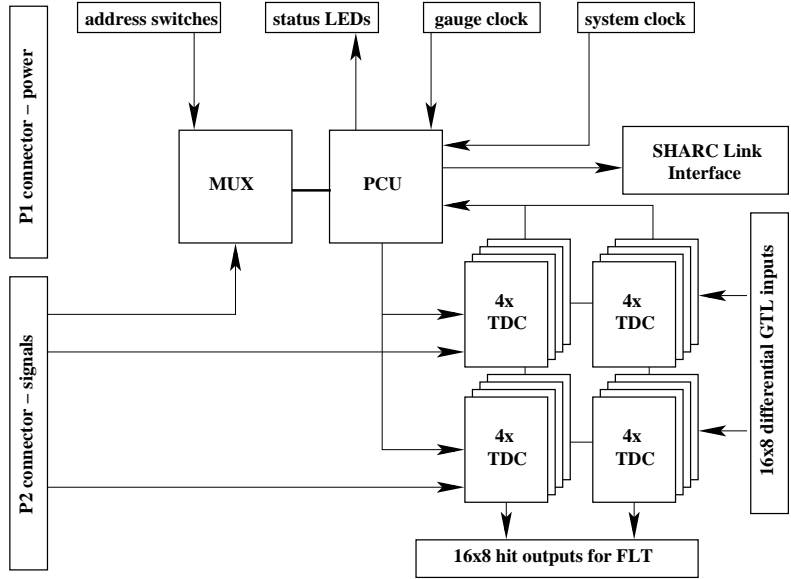}
    \caption{\label{board} Schematics of the TDC board structure for 
      time measurement.}
  \end{center}
\end{figure}

The TDC board (fig.\,\ref{board}), built in
9U VME format, comprises 16 TDC chips
 together with a DAQ interface (SHARC Link).
Except for the GTL inputs the signals on the board are TTL levels.
On the boards to be used as a hit register, an auxiliary chip converts
the differential amplifier outputs into single ended lines (`64 TTL
HIT INPUTS' in fig.\,\ref{chip}). All channels can be initialized in
parallel by a general Reset.

The main components on the TDC board are: 
\begin{itemize} 
\item[-] 16 TDC chips,
\item[-] SHARC Link interface,
\item[-] system clock + gauge clock (reference clock for calibration),
\item[-] Protocol Control Unit (PCU) + multiplexer (MUX),
\item[-] address switches (board and chip addresses).
\end{itemize}
The high-speed SHARC Link interface 
on the board (mainly fast drivers and transfer logic) transfers the
data for all channels from the FIFO on the TDC chip via its
point-to-point connection to a host system which, in HERA-B, is a
SHARC processor.
The 6-bit link port used for the transfer has four data lines, a clock
line and an acknowledge line. The link ports send  data  packed into 48 bit
words in direct communication with the SHARC memory. A fixed
clock/acknowledge handshake has been designed to control transfers
(only transmit cycles) to a SHARC-compatible receiver.

The Protocol Control Unit (PCU), which is a programmable logical
device (Lattice Semiconductors ispLSI 1048E), was employed to address
all TDC channels sequentially by using the Chip and Channel Select and
perform a packet-oriented protocol transmission consisting of
144~bytes. This Event Format Block Protocol is organized in three
major sections: the header field, the data field and the trailer field
(for details see \cite{zimmermann}).

The system clock determines the execution and transmission rates.  The
TDC board can utilize clock rates from 10~MHz to 30~MHz.  Thus the
maximal transfer rate to the SHARC board is 15~MByte/s (4 bit data
line at 30~MHz). In the Outer Tracker system a 27~MHz clock is
used.  A small clock circuitry generates a gauge clock rate of 10~MHz
to be used for the TDC online calibration.

The status of the TDC board is displayed by four LEDs indicating a
System Reset, a possible overflow situation, a Trigger Accept and a
fault condition. The overflow signal is a logical  OR  signal of all TDC
overflow flags. 
   
To limit noise on the ground plane the TDC board provides three
separate, independently grounded power systems: for the digital
control, for the TDC chips and for the FCS signals and the SHARC
outputs.  The user can select the optimal power scheme by bridges
connecting planes at several locations on board. For the HERA-B
application one common power supply turned out to be sufficient.  The
major environmental and electrical specifications of the board are
summarized in Table \ref{specs}.
  
\begin{table}
  \begin{center}
    \caption{\label{specs}Environmental and electrical specifications 
      of the TDC board.}
\vspace{1mm}
    \begin{tabular}{|lll|}\hline
      Operation temperature range&0 to 55 $^{\circ}$C&\\
      Operation humidity range&0 to 90 \%&  \\
      Height&366.8~mm&(9 VME HU)\\
      Depth&220.0~mm&\\
      Width&1.9~mm& \\
      Weight&736~g&\\\hline\hline
      {\bf Voltage}&{\bf Regulation}&{\bf Max. Current}\\\hline
      +5~V (digital)&$\pm$5\%&3.02~A\\
      +5~V TDC&$\pm$5\%&1.18~A\\\hline
    \end{tabular}
  \end{center}
\end{table}

\subsection{Calibration and Performance of the TDC system}
\label{tdcperform}               

The TDC system was designed to have a time measurement binning of
about 0.5~ns with a differential non-linearity of less than 3\%.  Due
to tolerances in the production process the actual value can differ
from the design value. The time bin size was measured for an
engineering sample of 20 TDC chips with a chip tester HP 82000
yielding  ($0.48~\pm~0.05$)\,ns at the working
temperature of 35~$^{\circ}$C with a temperature dependence of
($0.0015~\pm~0.0007$)\,ns/K. The quoted errors include the uncertainty
of the measurement device and the dispersion of the sample. For all
delivered charges of chips random test samples have verified that
these values remained stable. Because of the mapping of the required
maximal time interval onto 8 bits by the calibration procedure
described in section \ref{tdc_circuit}, the actual size of the time
measurement bin is uncritical. It has only to be assured that the
maximal time interval to be measured can be covered with 8 bits
(i.~e.~the time corresponding to the LSB of the TDC output cannot be
larger than time measurement bin).

All TDC boards were tested for linearity, dead time and time
resolution.  A dead time was measured to arise between sequential BX
clock cycles with typical values between 3\,ns and 5\,ns. This
decreases only slightly the maximally measurable time interval.

The {linearity} of the time measurement was verified for the whole
time range of 100\,ns.  A straight line fitted to the measured TDC
values over a range of 100~ns yields a slope of
($2.560~\pm~0.002$)\,counts/ns reflecting the mapping of the time
range onto 256 bins. The error is determined from fitting time
measurements of a single channel.

The resolution for a single TDC channel was measured to be 0.14~ns (1
standard deviation),
in agreement with the statistically expected value for a time
measurement in steps of 0.48~ns.
The corresponding uncertainty of the hit position of 11\,$\mu$m
contributes negligibly to the position resolution of the detector. The
long-term stability of the resolution 
is assured by the periodic calibration every 0.8\,s. Variations on a
shorter time scale could be caused by ripples from the TDC power
supplies. The measured resolution confirms that the chosen power
supplies are appropriate.

More details on the calibration of the TDC channels and the
performance of the TDC system can be found in \cite{zimmermann}.

\section{Electronics Installation and Commissioning}
\label{secinstall}

\subsection{Installation and System Integration}

The ASD-8 boards of the different categories were installed such that
for the innermost detector regions with the highest occupancies
amplifiers with a large signal-noise difference ($> 550\,$mV) were
chosen. The outer detector parts, contributing less to the acceptance,
were equipped with boards with a small signal-noise difference.
The CAN bus controlled distribution boards allow to set individual 
thresholds for groups of 8 ASD-8 boards (see section \ref{secpdb}).

\begin{figure}
\begin{center} \leavevmode
\epsfxsize=8cm  \epsfbox{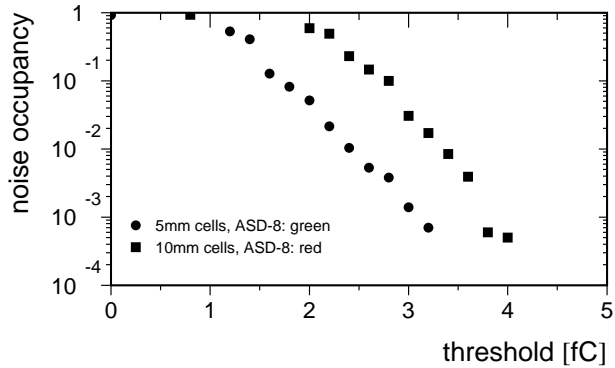}%
\end{center} 
\caption{Noise occupancies for two different sectors of a superlayer
after installation in HERA-B: The 5\,mm cells are equipped with ASD-8
boards from the best noise category, labelled as ``green'' (circles),
while the 10~mm cells are connected to boards of the worst noise
category, labelled as ``red'' (boxes).}
\label{fignoise}
\end{figure}  

The grounding and shielding scheme of the front-end electronics has
been implemented as described in section \ref{sec_ground}. After the
initial installation some rework had to be done to improve the system
stability.  For example, bad ground connections between either
amplifier and gas box or amplifier and cable shielding led to
oscillations of the amplifiers with their characteristic frequencies
of 40 to 80~MHz.

Another problem was a high noise level generated by the signal
connections between the TDC boards and the trigger link boards which
prepare the tracker hits for the First Level Trigger. This forced an
increase of the thresholds for the superlayers providing the trigger
input. To overcome this problem, high frequency filters were added and
the current transfer over these lines was reduced by removing the line
driver chips from the affected TDC boards. In this way the thresholds
could be set to similar values as in the other superlayers.

\subsection{Commissioning and Performance} 

The described front-end electronics has been used in HERA-B since
1997, the beginning of prototype tests, and since January 2000 all tracker
channels were fully equipped. They were running until the end of data
taking of the HERA-B experiment in 2003 with a very low failure rate,
with the exception of intermediate problems with the HV boards as
described below. For example, only about 0.1\% of the TDC
 boards and even less of the ASD-8 boards had to be replaced.

The noise occupancies (probability to find  a noise hit in
the readout time window) for two sectors of a superlayer after
installation in HERA-B are shown in fig.\,\ref{fignoise}.  The plot
shows a sector with 5 mm cells equipped with ASD-8 boards with low
noise thresholds and a sector with 10 mm cells and ASD-8 boards from
the worst noise class.  For the best ASD-8 boards the threshold can be
set as low as 2~fC, while for the worst category the threshold has to
be increased to about 3~fC to achieve a similar noise occupancy around
1\,\%. 

In the beginning of 2000 the HERA-B detector was 
completed and took first data from April to August. Unfortunately the
expected performance of the Outer Tracker system could not be reached
during this run. The most severe problem was a continuous loss
of channels due to high voltage breakdowns until finally  15\% of the
channels were lost. 

During the shutdown of HERA from autumn 2000 to summer 2001 major
repair and improvement work on the Outer Tracker had solved all
remaining problems. In particular, most of the high voltage breakdowns
could be identified to be due to two specific capacitors on the HV
board. These two were soldered with a different technique than the
rest, since they were placed on the opposite side of the HV boards
(section \ref{sec_hv_routing}). While 15 capacitors were glued prior
to the soldering process, the remaining two were positioned directly
on the board and soldered. It seems that the remaining soldering paste
under these two capacitors increased the probability of high voltage
breakdowns across them.  About 12000 of these capacitors had to be
replaced. This required access to every module and hence a complete
disassembly of the detector which was quite manpower and time
consuming.

After these repairs the Outer Tracker reached its final status. In the
data taking period 2002/2003 the detector was operated without any
specific problems. The design goals of the readout electronics have
been mostly reached. A slight deterioration of performance was
introduced by the necessity to raise the thresholds in some parts of
the detector as explained above. This deficit is mainly caused by the
relatively large channel-to-channel variations of the ASD-8
sensitivity and noise performance and could probably have been
overcome by applying individual thresholds for each channel. In
retrospective, the drawbacks of the grouping of thresholds seem to
preponderate the advantages of simplicity.

It was demonstrated that the electronics can handle high rates, up to
the design value of 40 MHz interaction rate with an occupancy in the
hottest channels of more than 20\%. However, finally the experiment
was mostly run at rates not exceeding 5~MHz.

The analysis of the 2002/2003 data yielded an overall good performance
of the Outer Tracker, with a high tracking efficiency of 96\% and a
track hit resolution of 370\,$\mu$m for tracks above 5\,GeV. The
resolution is much worse than the design value. This is partly due
to the not optimal threshold settings, as described above, but also to
deficiencies in the calibration and alignment procedures. Details of the
Outer Tracker performance are described in a separate paper
\cite{otr_perf}.

\section{Summary}

In this paper we have described the front-end readout system for the
112\,674 drift chamber channels of the HERA-B Outer Tracker
detector. The basic components are the amplifier-shaper-discriminator
chip ASD-8 and a customized TDC chip which provide the required high
integration density, low noise, high sensitivity, rate tolerance, and
low per-channel cost.
 
The high sensitivity of the ASD-8 amplifiers together with a large
chip-to-chip variation of the thresholds was a major challenge for the
implementation of the amplifiers. Grouping the chips according to
threshold and noise categories an economic system of threshold setting
was developed leading to an acceptable noise performance for the whole
system. An improvement is still possible by an individual
adjustment of the thresholds for each channel.

The TDC system is based on an ASIC 
which digitizes times in bins of about 0.5\,ns within a full scale of 256
bins. The time measurement is very stable due to an internal automatic
calibration procedure. In HERA-B the drift times are measured within
every bunch crossing period of 96~ns with respect to the external HERA
clock.  An integrated pipeline stores data for 128 events
satisfying the requirement for a dead-time free trigger and data
acquisition system at the design trigger rate.

The prototype tests and the analysis of data taken with the full
detector show that the front-end electronics of the Outer Tracker
fulfills the requirements posed on the detector for running in a high
rate environment.

\section*{Acknowledgements}
We thank our colleagues of the HERA-B Collaboration who made in a
common effort the running of the detector possible. The HERA-B
experiment would not have been possible without the enormous effort
and commitment of our technical and administrative staff. It is a
pleasure to thank all of them. 
 
We are grateful to Mitchell Newcomer for discussions and many
technical advices and to Karl-Tasso Kn\"opfle for carefully reading
the manuscript and for useful comments.

We express our gratitude to the DESY laboratory for the strong support
in setting up and running the HERA-B experiment. We are also indebted
to the DESY accelerator group for the continuous efforts to provide
good beam conditions.




\end{document}